\documentclass[sigplan]{acmart}
\usepackage{footnote}
\usepackage{xcolor}
\usepackage{tikz}

\setcopyright{none}
\renewcommand\footnotetextcopyrightpermission[1]{}

\copyrightyear{2024}
\acmYear{2024}
\setcopyright{rightsretained}
\acmConference[ICPP Workshops '24]{The 53rd International Conference on Parallel Processing Workshops}{August 12--15, 2024}{Gotland, Sweden}
\acmBooktitle{The 53rd International Conference on Parallel Processing Workshops (ICPP Workshops '24), August 12--15, 2024, Gotland, Sweden}
\acmPrice{}
\acmDOI{10.1145/3677333.3678273}
\acmISBN{979-8-4007-1802-1/24/08}

\newcommand\copyrighttext{%
  \footnotesize \textcopyright David Kacs 2024. This is the author's version of the work. It is posted here for your personal use. Not for redistribution. The definitive version was published in Workshop Proceedings of the 53rd International Conference on Parallel Processing, https://doi.org/10.1145/3677333.3678273}
\newcommand\copyrightnotice{%
\begin{tikzpicture}[remember picture,overlay]
\node[anchor=south,yshift=10pt] at (current page.south) {\fbox{\parbox{\dimexpr\textwidth-\fboxsep-\fboxrule\relax}{\copyrighttext}}};
\end{tikzpicture}%
}

\begin{document}

\title{Implementing OpenMP for Zig to enable its use in HPC context}

\author{Davids Kacs}
\email{D.Kacs@sms.ed.ac.uk}
\orcid{0009-0003-7387-6169}
\affiliation{%
  \institution{EPCC, University of Edinburgh}
  \city{Edinburgh}
  \country{United Kingdom}
}

\author{Nick Brown}
\orcid{0000-0003-2925-7275}
\affiliation{%
  \institution{EPCC, The University of Edinburgh}
  \city{Edinburgh}
  \country{United Kingdom}}

\author{Joseph Lee}
\orcid{0000-0002-1648-2740}
\affiliation{%
  \institution{EPCC, The University of Edinburgh}
  \city{Edinburgh}
  \country{United Kingdom}}

%%
%% The abstract is a short summary of the work to be presented in the
%% article.
\begin{abstract}

This extended abstract explores supporting OpenMP in the Zig programming
language. Whilst, C and Fortran are currently the main languages used to implement HPC
applications, Zig provides a similar level of performance complimented with several modern language features, such as enforcing memory safety. However, Zig lacks support for OpenMP which is the de facto threaded programming technology.

%The described implementation uses preprocessing techniques to augment Zig
%applications with the necessary calls to the OpenMP runtime to enable parallel
%execution.

Leveraging Zig's LLVM compiler tooling, we have added partial support for OpenMP to the Zig compiler and demonstrated that the performance attained by using Zig with OpenMP is comparable to, and in come cases exceeds, that of conventional HPC languages. Consequently we demonstrate that Zig is a viable and important programming technology to use for HPC, and this work paves the way for more HPC features to be added to Zig, ultimately providing HPC developers with the option of using a safer, more modern language for creating high performance applications.

\end{abstract}

%%
%% Keywords. The author(s) should pick words that accurately describe
%% the work being presented. Separate the keywords with commas.

\maketitle
\pagestyle{plain}
\copyrightnotice

% \pagebreak

\section{Introduction and Background} 

Developing High Performance Computing (HPC) applications is often a challenging and complex endeavour. These challenges most frequently arise from the fact that whilst the programming languages used by the community, C and Fortran 90, enjoy a long heritage in HPC, they lack modern language features. Whilst performance is often given as the reason for selecting such languages, it is our thesis that one can improve and simplify the process of developing HPC applications by enabling the use of more modern languages such as Zig\cite{zig}.

Zig is a systems programming language intended as a safer alternative to C. By providing a stronger type system and several optional runtime safety features\cite{kelly-zig}, such as array bounds checking, Zig programmers can avoid a large class of difficult-to-debug code issues relating to memory, whilst retaining performance comparable to that of C and Fortran due to both the language design and the LLVM\cite{llvm-paper} compiler implementation.

However, Zig was not designed for use in developing HPC codes, and as such it does not provide support for common compiler based frameworks that are ubiquitous in HPC such as OpenMP\cite{openmp}. Transitioning HPC software development from C to Zig offers enhanced safety and can significantly improve developer experience. However the lack of OpenMP support is a major blocker for adoption in HPC. 

In this extended abstract we report partial support for OpenMP in Zig, supporting the major features of OpenMP, and by reimplementing several kernels of NASA's NAS parallel benchmark suite\cite{npb} in Zig we demonstrate that multi-threaded Zig performance is comparable to more traditional HPC languages. Furthermore, we establish interoperability between Zig and Fortran, which is a crucial feature so that developers can leverage Zig as part of a wider legacy HPC application.

\section{Methodology}

OpenMP functionality was implemented in the Zig compiler by performing preprocessing during the early stages of compilation to identify and extract directives. This method was chosen as it enables calls to the OpenMP runtime to be inserted prior to invoking Zig's compile time execution engine and hence avoided making extensive modifications deep in the Zig compiler itself, limiting changes to 2651 additions and 30 deletions across 21 files.

Propagating the changes to the Internal Representation (IR) through the compiler was a key challenge and this motivated the decision to leverage preprocessing rather than directly manipulating the compiler's representation of the program. Whilst this approach was successful, the downside is that it does limit what type information is available during preprocessing. This limitation was overcome by leveraging Zig's generic programming features.

 \begin{figure}[h]
   \centering
   \includegraphics[width=\linewidth]{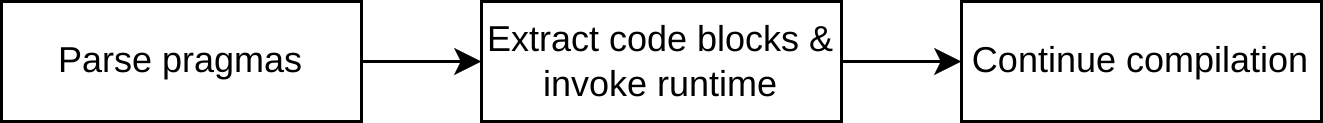}
   \caption{Overview of the process of intercepting and replacing OpenMP pragmas
   in the Zig compiler.}
   \Description{}
   \label{fig:pipeline}
 \end{figure}

As illustrated in Figure \ref{fig:pipeline}, integrating OpenMP into the Zig compiler involved two distinct parts, the parsing of pragmas and extraction of code blocks. Due to the minimal nature of the Zig language, it does not support any form of pragmas natively and is unlikely to do so in the future. As such they had to be added as comments, which is similar to OpenMP in Fortran. The compiler's existing parsing infrastructure was used to identify these tokens and integrate them into the compiler's program IR.

After parsing the OpenMP directives the compiler generates the code corresponding to all \verb|parallel| OpenMP regions. This is implemented by extracting the blocks of code identified by the \verb|omp parallel| pragmas into functions and passing pointers to these functions, and pointers to variables used in these functions, from the surrounding environment to the OpenMP runtime library. Similarly, OpenMP's worksharing loop directive was preprocessed and we add a runtime library routine call to calculate the loop bounds.

In addition to the two directives described above, support for many of the OpenMP clauses was also implemented. This includes support for the data sharing clauses \verb|shared|, \verb|private|, and \verb|firstprivate|, the \verb|schedule| clause which determines worksharing loop distribution over threads, and the \verb|reduction| clause which reduces values across loop iterations in a thread safe manner. We link against the LLVM projects' OpenMP runtime library \cite{llvmRuntime} in this work. 

\section{Results} \label{sec:results}

\subsection{Benchmarking}

We aim to demonstrate the benefit of Zig with OpenMP using NASA's NAS parallel benchmark suite (NPB), which provides individual kernels that are representative of CFD applications. All benchmarks leverage the parallel region and worksharing loop directives, with the set of clauses varying between benchmark. Results, including total runtime and speedup, were compared to the reference implementation, with speedup calculated relative to single-thread execution.

%In order to ensure the benchmarks were compared in a fair manner, the same implementation of OpenMP was used for all three languages, necessitating the use of the Clang and Flang LLVM-based compilers for C and Fortran benchmarks respectively. All benchmarks were ran on up to 128 cores on a single node of the ARCHER2\cite{archer2} system.

One challenge we faced is that the Zig compiler was not able to integrate with Fortran. We developed an approach where invoking Fortran procedures from Zig was possible by declaring these as C linkage functions using pointer arguments, and appending underscores to function names to comply with the Fortran compiler's name mangling scheme.

Table \ref{tab:results} reports the runtime for the Conjugate Gradient (CG), Embarrassingly Parallel (EP) and Integer Sort (IS) NPB benchmarks (class C problem size), along with a  Mandelbrot Set benchmark. The runtime of OpenMP reference implementations (CG and EP benchmarks are in Fortran, whereas IS and the Mandelbrot Set in C) is reported by \emph{Reference} and our approach in Zig by \emph{Zig+OpenMP}. All runs where over 128 cores which is a single node of the ARCHER2 supercomputer.

\begin{table}[h]
  \caption{Performance of benchmark reference implementation against our Zig with OpenMP approach running over a single node (128 cores) of ARCHER2 an HPE Cray EX}
  \label{tab:results}
\begin{center}
\begin{tabular}{|c|cccc| } 
 \hline
 \textbf{Version} & \multicolumn{4}{c|}{\textbf{Runtime (s)}} \\
  & CG & EP & IS & Mandlebrot\\
 \hline
 Reference & 2.07 & 1.42 & 0.24 & 5.08\\ 
 Zig+OpenMP & 1.81 & 1.27 & 0.27 & 5.36\\ 
 \hline
\end{tabular}
\end{center}
\end{table}

It can be seen that across all these benchmarks our Zig implementation provides performance which is comparable to the reference implementations. When compared against Fortran with OpenMP, the Zig implementation achieved approximately an 11\% performance improvement for EP and 12\% for CG benchmarks. The C with OpenMP reference implementations slightly outperformed Zig, 11\% for IS and 5\% for Mandlebrot.

\section{Conclusion}

This work demonstrates the feasibility and benefits of implementing OpenMP in
Zig, enhancing its suitability for HPC applications. Future work could further
explore Zig's HPC potential, including developing native Zig benchmarks and
improving interoperability with Fortran.

%%
%% The next two lines define the bibliography style to be used, and
%% the bibliography file.
\bibliographystyle{ACM-Reference-Format}
\bibliography{base}

\end{document}